\documentclass{article}

\usepackage[preprint]{neurips_2024}
\usepackage{amsmath}
\usepackage{graphicx}

\usepackage[utf8]{inputenc} 
\usepackage[T1]{fontenc}    
\usepackage{hyperref}       
\usepackage{url}            
\usepackage{booktabs}       
\usepackage{amsfonts}       
\usepackage{nicefrac}       
\usepackage{microtype}      
\usepackage{xcolor}         

\title{Use-dependent Biases as Optimal Action under Information Bottleneck}

\author{%
  Hokin X. Deng\\
  Krieger School of Arts and Sciences\\
  Johns Hopkins University\\
  Baltimore, MD 21218 \\
  \texttt{xdeng13@jhu.edu} \\
  \And
  Adrian M. Haith\\
  Department of Neurology\\
  Johns Hopkins School of Medicine\\
  Baltimore, MD 21205 \\
  \texttt{adrian.haith@jhu.edu} \\
}

\begin{document}

\maketitle

\bibliographystyle{apalike}

\begin{abstract}
    Use-dependent bias is a phenomenon in human sensorimotor behavior whereby movements become biased towards previously repeated actions \citep{tsay_dissociable_2022,verstynen_how_2011}. 
  Despite being well-documented, the reason why this phenomenon occurs is not yet clearly understood. Here, we propose that use-dependent biases can be understood as a rational strategy for movement under limitations on the capacity to process sensory information to guide motor output. We adopt an information-theoretic approach to characterize sensorimotor information processing and determine how behavior should be optimized given limitations to this capacity.
  We show that this theory naturally predicts the existence of use-dependent biases. Our framework also generates two further predictions. The first prediction relates to handedness. The dominant hand is associated with enhanced dexterity and reduced movement variability compared to the non-dominant hand, which we propose relates to a greater capacity for information processing in regions that control movement of the dominant hand. Consequently, the dominant hand should exhibit smaller use-dependent biases compared to the non-dominant hand. The second prediction relates to how use-dependent biases are affected by movement speed. When moving faster, it is more challenging to correct for initial movement errors online during the movement. This should exacerbate costs associated with initial directional error and, according to our theory, reduce the extent of use-dependent biases compared to slower movements, and vice versa. We show that these two empirical predictions, the handedness effect and the speed-dependent effect, are confirmed by experimental data. 
\end{abstract}

\section{Introduction}

\paragraph{} It has been widely observed that when people repeatedly perform the same actions, their subsequent movements become biased toward those actions \citep{tsay_dissociable_2022,verstynen_how_2011}. Known as use-dependent biases, this sensorimotor phenomenon has been shown to be robust across various domains: they persist in both kinematic and force spaces \citep{diedrichsen2010use, marinovic_action_2017}, and across different motor effectors \citep{wood2021consistency,tays2020consolidation, seegelke2021repetition}. Nevertheless, it remains unclear why humans have these sensorimotor biases and whether they might confer any benefit for control.  Here, we propose that use-dependent biases can be understood naturally as a consequence of information bottlenecks in sensorimotor control.

\paragraph{} It is widely recognized that our ability to sense and act in the world is limited by inherent physical and biological constraints on information processing and computation \citep{bryant2023physical, zenon2019information}. This idea has been extensively explored in cognitive science in the context of decision making between discrete action choices \citep{lieder_resource-rational_2020,lai_human_2024,lai_policy_2021}. However, the implications of limited information processing has hardly been considered in the context of sensorimotor behaviors for continuous action spaces \citep{kording_bayesian_2006,todorov_optimal_2002}. Here, we adopt an information-theoretic approach to quantifying information processing requirements for action selection \citep{tishby2010information, lai_policy_2021} and show that this framework naturally predicts the occurrence of use-dependent biases. We further extend this framework to generate two new predictions associated with use-dependent biases and handedness, and the effect of movement speed on use-dependent biases.

\paragraph{} Handedness generally refers to the the inclination to favor one particular hand to perform tasks and is a ubiquitous trait in human populations \citep{annett_distribution_1972, mcmanus_half_2019}. A key feature of handedness is that the non-dominant hand has greater movement variability compared to the dominant hand \citep{roy_manual_1986, salimpour_motor_2014}. We argue that this increased movement variability can be understood as reflecting more limited information processing capacity in networks responsible for controlling the non-dominant hand. According to our theory, this reduced information capacity should, in turn, lead to exacerbated use-dependent biases in the non-dominant hand.

\paragraph{} Our theory also predicts a speed-dependent effect in use-dependent biases. At slow movement speeds, the initial direction of movement does not matter that much as it is easy to make online feedback corrections during movement. As movement speed increases, however, it becomes more and more difficult to make such corrections before the end of the movement. Therefore, the cost of one's initial movement direction deviating from the target angle rises more sharply when moving quickly compared to moving slowly \citep{haith2015hedging}. Our theory predicts that the extent of use-dependent biases are sensitive to these costs and, thus, use-dependent biases should be reduced for fast compared to slow movements.

\paragraph{} We validated both of these two predictions. For the handedness effect, we conducted a new experiment to compare the extent of use-dependent biasing effects in the dominant and non-dominant hands. For the speed-dependent effect, we re-analyzed data from experiments in a previous study \citep{wong_motor_2017}. Limited information processing capacity thus provides a natural and parsimonious explanation for a range of phenomena related to use-dependent biases in movement, supporting the view that information processing constraints play an important role in shaping human motor behavior.

\begin{figure}[h]
  \centering
  \includegraphics[width = 13cm]{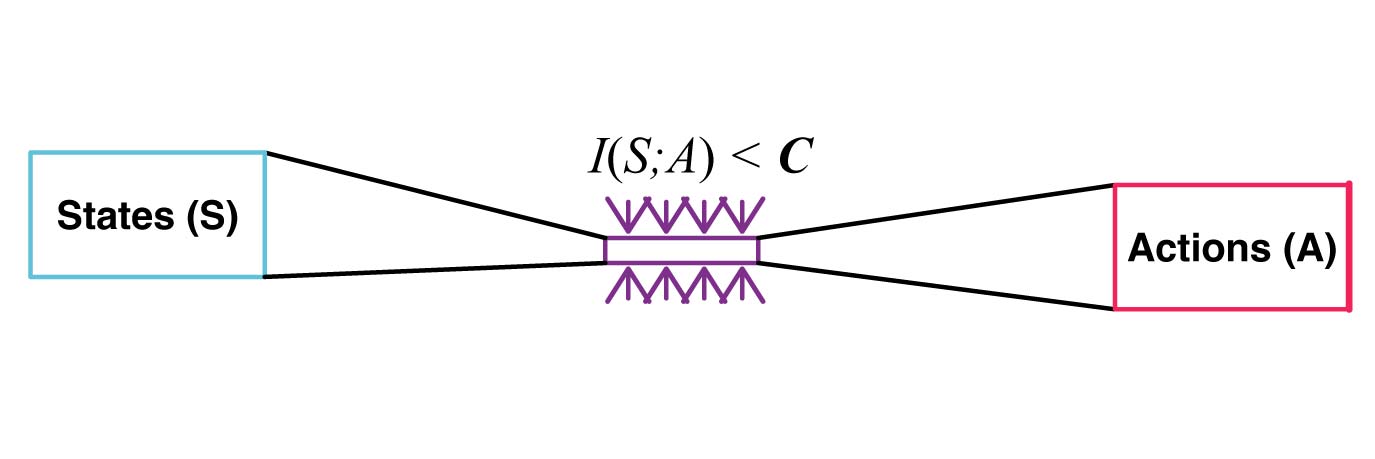}
  \caption{Illustration of an information bottleneck in sensorimotor control. The sensorimotor system can be considered, in abstract terms as an information channel transforming sensory states into motor actions. Limited computational resources of the sensorimotor system can be understood as an information bottleneck between sensory states and motor actions. Mathematically speaking, the mutual information $I(S;A)$ between states $S$ and actions $A$ must not exceed some bound $C$. }
\end{figure}

\section{Theory}

\subsection{General Framework}

\paragraph{} To characterize information processing in sensorimotor control and examine the implications of limited information processing, we adopt an information-theoretic framework which posits that information processing costs for a given behavior can be quantified through the mutual information $I(S;A)$ between sensory states of the task $S$ and motor actions $A$ \citep{tishby2010information}. 

\paragraph{} Intuitively, the mutual information term describes how finely tuned actions are to the current state. Performing the same action regardless of one's current state requires no sensorimotor information processing. By contrast, if different actions are reliably selected in distinct states, this requires information processing. The more finely different states are distinguished in order to generate different actions, the more information processing is required, reflected in a greater mutual information between states and actions. For example, a novice tennis player will likely swing at an incoming ball in the same way, regardless of the type of spin on the ball, whereas an expert player will factor the ball’s spin into how they strike the ball. The expert player's action selection requires more extensive processing of the incoming sensory information and is thus associated with a greater information processing capacity. 

\paragraph{} Limitations on information processing can be conceptualized as an upper bound on the possible mutual information between states and actions \citep{tishby2010information,lai_policy_2021}. Formally, suppose we have a sensorimotor task in which we wish to select actions to minimize the expected value of some cost $J(s,a)$ we can consider a policy that optimizes this cost subject to the constraint that the mutual information between states and actions must not exceed some capacity limit $C$, i.e. 

\begin{equation}
    \min E [ J(s, a)] \quad \textrm{s.t.} \quad I(S;A) \leq C.
    \label{eq:one}
\end{equation}

The solution to this constrained optimization problem takes the form of a stochastic policy $p(a|s)$ that determines a probability distribution over actions for any given task state . This optimal distribution can be shown to be of the form (Appendix \ref{pro:derivation}):

\begin{equation}
    p(a|s) \propto p(a) e^{- \beta J(s,a)}.
    \label{eq:two}
\end{equation}

This expression comprises two factors. The rightmost factor $e^{- \beta J(s,a)}$ states that less costly actions should be selected with greater probability, analogous to softmax action selection. Here $\beta$ is a parameter that depends on the information capacity $C$ and acts as a temperature parameter, i.e. governing the extent to which this distribution is concentrated on the least costly actions. The second factor $p(a)$ serves to bias this distribution towards the overall distribution of actions, independent of the current task state. This equation immediately predicts the well-known phenomenon in motor control of use-dependent learning, whereby one’s actions are biased towards one’s recent history of actions \citep{marinovic_action_2017,verstynen_how_2011,tsay_dissociable_2022}. Suppose that the state $s$ corresponds to the direction of a target to be reached to, while the action $a$ corresponds to the direction of the reach. If, as is typical in theories of motor control \citep{haith2015hedging}, one assumes a simple quadratic cost function $J = \frac{1}{2}(s-a)^2$, and that $p(a)$ follows a Gaussian distribution $p(a) \sim \mathcal{N}(s_0, \frac{1}{\epsilon})$ (which here is parameterized by precision $ \epsilon$ rather than variance, for convenience), the optimal policy under an information bottleneck becomes 

\begin{equation}
    p(a|s) \propto e^{ -\frac{1}{2} \left(a - \frac{\beta s + \epsilon s_0}{ \epsilon +\beta}\right)^2}.
    \label{eq:thr}
\end{equation}

This represents a Gaussian distribution who’s mean is shifted away from the target distribution and towards the repeated action $s_0$. This equation exactly describes the phenomenon of use-dependent learning, in which the biases towards a previously repeated movement is found to be proportional to the distance between the repeated target and the new target, and is also sensitive to the variance of movements around the repeated direction \citep{verstynen_how_2011,marinovic_action_2017,tsay_dissociable_2022}. Notably, according to this equation, the extent of the bias also depends on the bound on the mutual information between states and actions, through the parameter $\beta$.

\subsection{Handedness Effect}

\paragraph{} Human handedness is typically defined as a preference for which hand to use. However, handedness is strongly associated with control performance. In particular, variability of movement has consistently been shown to be greater in the non-dominant hand compared to the dominant hand.

\paragraph{} We propose that this increased movement variability in the non-dominant hand could be understood in terms of a reduced information capacity for controlling the non-dominant hand. That is, we propose that the information capacity $C$ for selecting actions for the dominant (D) and non-dominant (ND) hands might be different, with $C_{ND} < C_D$. Since the relationship between $C$ and $\beta$ is known to be monotonic \citep{lai_policy_2021}, this implies that the values of $\beta$ will also be different, i.e.
\begin{equation}
    \beta_{ND} < \beta_D
\end{equation}
and will lead to greater variability of $p(a|s)$ for the non-dominant hand compared to the dominant hand.

\paragraph{} These different values of $\beta$ will also, according to Equation \ref{eq:thr}, lead to differing shifts in the mean of $p(a|s)$, such that the non-dominant hand should exhibit a greater use-dependent learning effect compared to the dominant hand.

\subsection{Speed-dependent Effect}

\paragraph{} When moving faster, it becomes more difficult to correct for initial directional errors during the course of the movement because the time to reach the target is much shorter. As such, the cost of the initial direction of movement deviating from the target angle should increase more steeply when when moving quickly compared with moving slowly.

\paragraph{} This intuition can be more formally justified using optimal control theory \citep{haith2015hedging, liu2007evidence, haith2013theoretical}. Following Liu and Todorov \citep{liu2007evidence}, we modeled control of the hand during a reaching movement as a discrete-time linear dynamical system with state $x_t$, and subject to time-varying controls $u_t$: 

\begin{equation}
    x_{t+1} = A_{d}x_{t} + B_{d}u_{t}
    \label{eq:fou}
\end{equation}

and assume that the hand is controlled to minimize a terminal cost 

\begin{equation}
    J_T= J_{x} + J_u
    \label{eq:fiv}
\end{equation}
where we assume that $Jx$ is a quadratic function penalizes endpoint deviations from the goal:

\begin{equation}
    J_{x} = \frac{1}{2}(x_{T} - s)^2
    \label{eq:six}
\end{equation}

while $J_u$ is an effort cost that is a quadratic function of the overall sequence of motor commands: 

\begin{equation}
    J_{u} = w_u \sum_{t=1}^T u^2_t
    \label{eq:sev}
\end{equation}
where $w_u$ describes the relative cost associated with effort versus accuracy.

\paragraph{} To model use-dependent learning, we are primarily interested in the costs associated with  deviating from the correct target heading direction at the outset of the movement, which is characterized by the \emph{cost-to-go} $V_t(x,s)$ at the start of the movement, which will be a quadratic function of the initial movement direction.

\paragraph{} We computed this cost-to-go function for the optimal feedback control model from \citep{liu2007evidence}, using a movement duration of either 250 ms (fast movements) or 500 ms (slow movements). We translated this cost-to-go function into our information-limited action selection model by quantifying the cost of the initial movement direction deviating from the target direction as the position component of the associated cost-to-go function at the start of the movement. We calculated that this decrease in movement duration lead to an approximately 7-fold increase in the cost associated with initial movement direction errors.

\paragraph{} This optimal feedback control model provided a rational estimate of the expected difference in initial movement direction costs for fast and slow movements which, when substituted into Equation \ref{eq:two} led to specific predictions of how movement speed should affect the magnitude of use-dependent biases. We generated quantitative predictions about the effect of movement speed on use-dependent biases by setting the value of $\beta$ in equation \ref{eq:two} to predict a use-dependent bias of approximately 3° for slow movements (consistent with data from \citep{wong_motor_2017}) and then used the expected increased initial direction costs to predict the magnitude of use-dependent biases for fast movements.

\begin{figure}[b]
\vspace{1 cm}
  \centering
  \includegraphics[width = 13cm]{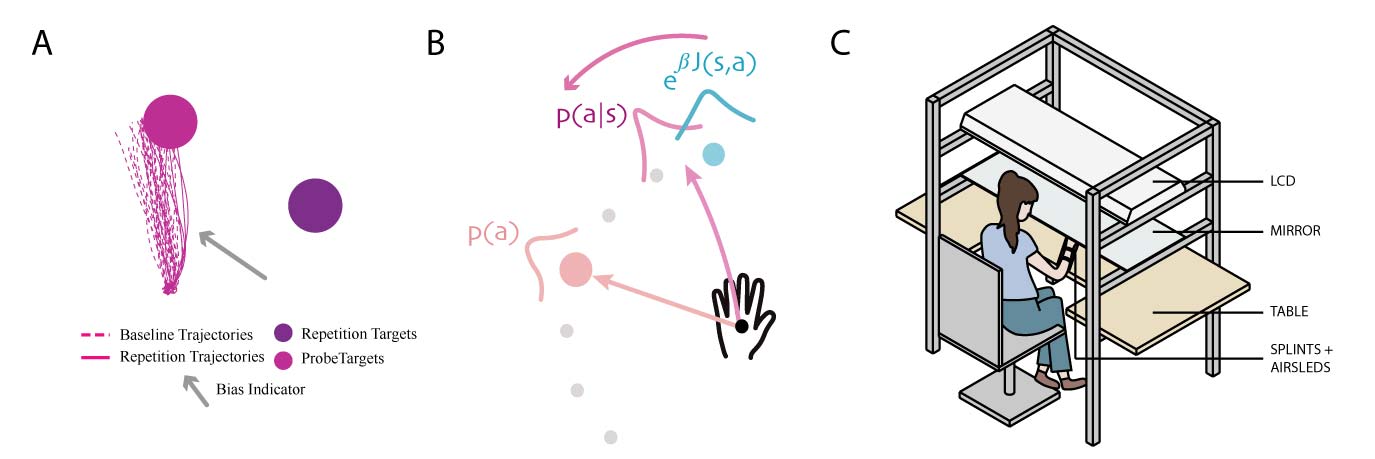}
  \caption{A. Use-dependent biases in reaching movements from a representative participant. B. If we assume that the cost is a quadratic function of reach error, and that the overall distribution of movement directions $p(a)$ is Gaussian, this theory predicts that actual movement direction will follow a Gaussian distribution whose mean is biased towards previously generated actions – exactly predicting the previously reported phenomenon of use-dependent biases. C. Illustration of experimental setup in which participants made planar arm movements while viewing a display. }
\end{figure}

\section{Experiment}

\subsection{Logistics and Equipment}

\paragraph{} All participants reported no prior history of neurological disorders. All methods were approved by the Johns Hopkins School of Medicine Institutional Review Board and were carried out in accordance with relevant guidelines and regulations. Written informed consent was obtained from all participants in the study. All participants received financial compensation for their participation.

\paragraph{} Participants were seated in front of a table with their hands supported on the table by frictionless air sleds. The positions of both hands of each participant were tracked at 130Hz with a Flock of Birds magnetic tracker (Ascension Technology, Shelburne, VT) placed under each index finger. Participants viewed stimuli on a horizontal mirror that reflected an image of an LCD monitor (60 Hz) and obscured vision of both hands (Figure 2a). The distance between the LCD screen and the mirror was 25 cm. The distance between the LCD screen and the air sled was 25 cm, so that the cursor’s image was veridically aligned with the tip of the index finger. Targets (diameter: 10 mm) and a hand-controlled blue cursor (diameter: 5 mm) were also presented through the same display.

\subsection{Handedness Effect Experiment}

\paragraph{} A total of 28 participants were recruited for this study (aged 19-28; 11 male). Participants performed a planar reaching task consisting of 14 blocks total. Participants completed 7 blocks with one hand before switching to the other hand to complete a further 7 blocks that followed the same structure. 

\paragraph{} For each hand, there were 7 possible targets evenly spaced along an arc, all 12 cm from a fixed, central start position on the body mid-line. For the right hand, the target locations were 10°, 30°, 50°, 70°, 90°, 110°, and 130°clockwise from the straight-ahead direction. This target configuration was mirrored for the left hand. The first two blocks for each hand were baseline blocks of 98 trials each, with each of the 7 targets appearing 14 times in a pseudo-random order. Participants then completed seven repetition blocks of 100 trials each. In these repetition blocks, the central target of the 7 targets appeared 70 times within the block, and each of the other 6 targets appeared 6 times each. All trials were interleaved in a pseudo-random order. Participants additionally received corrective feedback about the speed of their movements in the form of the target changing color if they moved too quickly (red) or slowly (blue), in order to maintain a consistent movement speed across hands.

\paragraph{} Within each trial, the participant moved the cursor inside a yellow start circle (60 mm diameter) that was positioned on the body mid-line. Once the cursor remained within the start circle for 50 ms, one of the seven targets appeared (red circle, 10 mm diameter). The participant then performed a “shooting” movement to guide the cursor through the target without having to stop inside it. Once the cursor left the start circle, it became no longer visible until it reached a distance of 12.5 cm away from the start circle. This was done to ensure that participants did not make feedback corrections during movement and to reduce the possibility of them using online feedback to correct directional biases from trial to trial. Participants then returned their hand to the start circle with the cursor visible. 

\subsection{Speed-dependent Effect Dataset}

\paragraph{} To test our hypothesis about the effect of speed on use-dependent biases, we reanalyzed a previously collected dataset from \citep{wong_motor_2017}. This dataset consistent of 16 participants who completed ten blocks of trials, grouped into two sets of five blocks. Each set was performed at either a ‘Fast’ or ‘Slow’ speed. A set consisted of one training block with only single-target trials to practice the required movement speed (48 trials), followed by four test blocks, each comprising 48 ‘single-target’ and 48 ‘dual-target’ trials (12 trials at each target-separation angle) randomly intermixed.

\paragraph{} In ‘dual-target’ trials, two targets were initially displayed concurrently, and participants were required to initiate a movement without full knowledge of which target was the correct one. We focused our analysis, however, on ‘single-trials’ in which only a single target was presented.

\paragraph{} In these single-target trials, one target appeared at one of eight possible locations: either \( +7.5^\circ \) and \( -7.5^\circ \), \( +15^\circ \) and \( -15^\circ \), \( +22.5^\circ \) and \( -22.5^\circ \), or \( +30^\circ \) and \( -30^\circ \). After a random interval (450-1000 ms), an auditory beep cued participants to initiate a movement through the target. A trial was considered successful if the cursor moved through the correct target circle while satisfying the velocity criterion for that block.

\subsection{Data Analysis}

Both datasets were analyzed in the same way. Data were smoothed to remove instrument noise using a 3-rd order Savitzky-Golay filter and numerically differentiated and smoothed again to obtain velocity. Movement initiation was determined based on the time at which movement velocity exceeded a threshold of 0.02 ms$^{-1}$. Initial movement direction was quantified as the direction of the velocity vector 100 ms after the time of movement initiation.

\section{Results}

\subsection{Handedness Effects on Use-Dependent Biases} 

\paragraph{} Previous work has shown that the size of use-dependent biasas increases linearly with distance from the repeated target \citep{verstynen_how_2011}. We therefore estimated a linear slope relating target angles to initial movement direction. We expected this slope to be close to $1$ in the baseline blocks, indicating no use-dependent biases. Critically, we predict that the slope would get smaller in repetition blocks, as use-dependent biases manifest. We calculated the magnitude of the use-dependent bias effect as the change in this slope in blocks with and without repetition.

\paragraph{} From our experimental data, repetitions of movements significantly reduced the slope for the non-dominant hand ($\Delta$ Slope non-dominant = - 0.072 ± 0.040, s.e.). However repetition had a smaller effect on the dominant hand ($\Delta$ Slope dominant = -0.039 ± 0.049, s.e.). We performed a repeated two-way ANOVA on the slope measure, which revealed a significant hand-by-condition interaction (p = 0.009, F = 7.92), confirming our prediction that use-dependent biases would be greater in the non-dominant hand compared to the dominant hand. We further performed a bootstrap analysis as an alternative approach to assess whether the size of the use-dependent learning effect was different across hands. We generated 5,000 synthetic (“bootstrapped”) datasets by resampling from our participants and computed the difference in the change in slope ( $\Delta$Slope non-dominant - $\Delta$Slope dominant) for each sample. Figure 2A shows the distribution of this statistic across samples. We determined a $99 \% $ confidence interval of [0.0037, 0.0639] 

\subsection{Effect of Movement Speed on Use-dependent Biases} 
\paragraph{} We re-examined the results of a previous study \citep{wong_motor_2017} in which movement speed was systematically varied within participants. These data showed a clear effect of movement speed on use-dependent biases (Figure 2B). On Slow movement trials, the reach direction was consistently biased towards the centre of the array of potential target locations (one-way ANOVA, F(3,15)=8.85, p<0.001). However, these use-dependent biases were absent in Fast movement trials. This effect was statistically significant (two-way ANOVA, interaction between speed eccentricity (F(3,15)=2.81, p=0.04; statistical analysis from \citep{wong_motor_2017}).

\begin{figure}[h]
  \centering
  \includegraphics[width = 13cm]{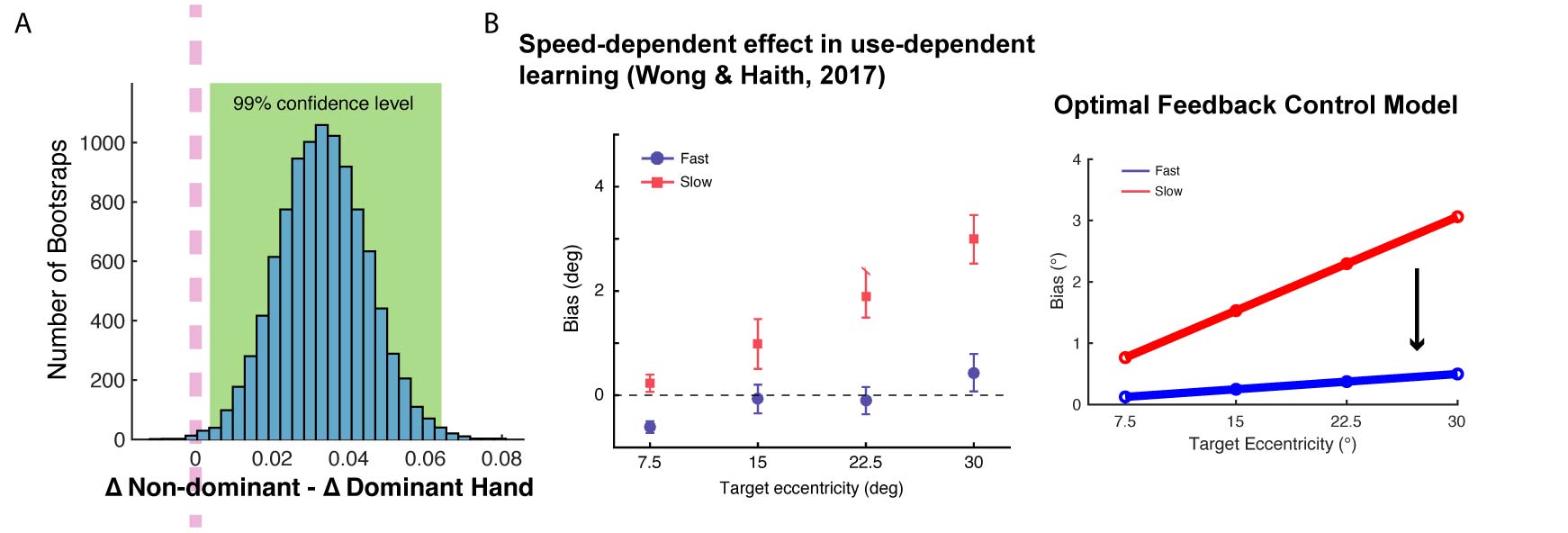}
  \caption{A. Bootstrap analysis: histogram plot of the magnitude of the use-dependent biases in 10,000 bootstrapped samples; green area indicates $99\%$ confidence interval of [0.0037, 0.0639] B. Use-dependent biases at Slow and Fast movement speeds. Left, data from \citep{wong_motor_2017} showing the effect of movement speed on the magnitude of use-dependent biases. Right, theory predictions. We fit the model to data for fast movements (red) and predicted the change in use-dependent biases when moving more quickly, based on increased directional error costs predicted by optimal control theory.}
  \label{fig:thr}
\end{figure}

We asked whether the direction and magnitude of this effect would be predicted by our theory. We set the parameters of our model ($\beta$ in Equation \ref{eq:thr}) so that it predicted a use-dependent bias of approximately 3° for a movement with 30° eccentricity, consistent with the data from \citep{wong_motor_2017}. We then used an optimal feedback control model \citep{liu2007evidence} to predict how much the initial direction cost would be increased for faster movements (half the duration to reach the target). This model suggested that the costs on initial directional error would increase by a factor of 6.73. When we applied this increased cost to predict the magnitude of use-dependent biases for fast movements, it predicted that the use-dependent biases would be virtually eliminated (Figure \ref{fig:thr}).

In short, the experimental data from this previous study confirmed the speed-dependent effect prediction in use-dependent biases from our theory. 

\section{Discussion}

\paragraph{} Traditional theories in sensorimotor control suggest that we optimize our movements to achieve the highest possible level of efficiency and precision, focusing on the end result of motor output without taking into consideration the underlying computational costs associated with selecting and generating these movements \citep{todorov_optimal_2002, harris_main_2006, lockhart_optimal_2007, wong_motor_2017}. These theories prioritize the outcome of actions, such as reaching a target or performing a movement with minimal error, and utilize models that emphasize the biomechanical and physiological constraints on motor performance \citep{flash_coordination_1985, seethapathi_metabolic_2015, seethapathi_step--step_2019}.
\paragraph{} The information bottleneck approach has been recently applied in the context of cognitive decision making as a general quantitative framework for understanding resource-rational computations \citep{lieder_resource-rational_2020, correa_humans_2023, cantlon_uniquely_2024}. Whereas some resource-rational theories seek to understand the impact of resource limitations within specific architectures for cognitive computations, e.g. the impact of limited working memory capacity \citep{hong_episodic_2024}, the information bottleneck perspective provides a more abstract framework that is more agnostic to specific underlying computational architectures and can be couched at a purely behavioral level. Within the domain of cognitive science, the information processing cost is generally understood as either approximating costs associated with cognitive operations \citep{parr_cognitive_2023}, or as expressing the need to compress learned policies for subsequent storage and retrieval from memory \citep{lai_human_2024}. 

\paragraph{} In the context of motor control, however, we suggest that the nature of underlying information processing costs might be quite different. Cognitive processes are believed to play a critical role planning and executing movements \citep{sayali_neural_2021} and thus computational constraints on cognition ought naturally to also apply to motor control that depends on cognition. However, continuous motor behaviors like point-to-point reaching or walking are known to be rapid and automatic \citep{seethapathi_step--step_2019, schubert_mechanism_2024}; that is, they do not appear to depend on cognitive resources or computations. In this case, information processing costs are likely primarily associated with the precision with which sensory information can be used to guide motor output, which may depend more on the volume of neural resources available to implement this computation, potentially to save energy \citep{laughlin_energy_2001, seethapathi_metabolic_2015}, rather than on the extent to which the policy can be compressed in long-term memory \citep{lai_human_2024}.
\paragraph{}We suspect that information-theoretic constraints on motor performance might have deep and wide-ranging implications for motor control. Here, we examined the most straightforward implication of this theory, showing that it naturally accounts for well-documented use-dependent learning effects  in reaching. It has previously been suggested that use-dependent learning could be understood in terms of Bayesian inference of the target location \citep{verstynen_how_2011, tsay_dissociable_2022}. Although the equations describing behavior that we derived from the information bottleneck theory are directly analogous to Bayesian inference, there are several important conceptual differences from a Bayesian account of use-dependent biases. Critically, the information bottleneck theory is analogous to Bayesian inference of the motor commands needed to achieve the task  \citep{todorov_optimal_2002, botvinick_planning_2012}, rather than of the location of the target of the reach. The information bottleneck theory goes further, however: whereas Bayesian theories frame action variability as something that must be factored into action planning, the information bottleneck theory posits that even the degree of action variability is a consequence of the information bottleneck – i.e. variability is a consequence of intrinsic constraints or costs associated with neural computation, rather than a consequence of extrinsic factors associated with peripheral mechanisms of movement execution. This perspective is in-keeping with findings suggesting that motor variability originates from the brain, rather than in the motor periphery \citep{jones_sources_2002, chaisanguanthum_motor_2014, churchland_central_2006}. Finally, the information-bottleneck perspective also predicts that the patterns of bias and variability are shaped by the cost structure of the task. We showed that speed-dependence of use-dependent movement biases are also naturally explained by the fact that moving faster alters the costs associated with directional error (due to the reduced ability to make online corrections when moving quickly). This phenomenon cannot be explained by more simplistic Bayesian theories. Therefore, unlike alternative perspectives, therefore, the information bottleneck theory provides a unified and parsimonious account of both the origins of movement variability and use-dependent biases.

\paragraph{}This information-theoretic perspective on motor control led us to propose a novel explanation for the phenomenon of human handedness, whereby handedness reflects differential information processing capacities between the dominant and non-dominant hands. According to our hypothesis, the dominant hand, typically used for more complex and skilled tasks, possesses a higher information processing capacity. This difference most obviously manifests as an increased precision of movement in the non-dominant hand, which is predicted by the theory and is well-established in the literature \citep{woodworth_accuracy_1899, elliott_asymmetries_1993, elliott_influence_1994, takagi_command_2022}. More generally, the increased information capacity may also allow for a richer movement repertoire and enhanced dexterity. The implications of reduced information capacity for more dynamic control tasks is challenging to determine, but we expect future work to determine whether our theory might also account for observed asymmetries in control strategy across the two hands \citep{sainburg_differences_2000, sainburg_evidence_2002, sainburg_handedness_2005, sainburg_convergent_2014}. 

\paragraph{}Our information-theoretic perspective also raises the possibility of a normative explanation for handedness. Current theories of the origin of handedness tend to be focus on the mechanisms that dictate handedness (e.g. brain lateralization \citep{sha_handedness_2021}), without necessarily addressing why it would make sense for there to be an asymmetry across hands. Though our current theory is not normative in this sense, it could in principle be extended to allow for a limited overall information processing capacity budget that must be divided across the two hands. The optimal allocation of information processing capacity, or ‘bandwidth’ between the hands would ultimately depend on the statistics of tasks that could be accomplished using either just one hand (in which case asymmetry is good) or require dexterous coordination of both hands (in which symmetric information capacity would be preferred). Such a theory could provide an explanation of where the asymmetry of information processing bandwidth arises from in individuals and/or species.

\newpage

\bibliography{neurips_2024}

\newpage

\appendix

\section{Optimal Action under Information Bottleneck}
\label{pro:derivation}
\paragraph{} Here we model our sensorimotor policy as a stochastic function $p(a|s)$, where the action $a$ (reaching direction) is selected based on the state $s$ (target direction). In canonical optimal sensorimotor control theories \citep{todorov_optimal_2002, haith2015hedging}, goal-directed human reaching could be modeled as finding the stochastic policy $p(a|s)$ which minimizes an accuracy cost that penalizes the distance between actual arm direction and target direction. This is typically given as,
\begin{align*}
    \min_{p(a|s)} \mathcal{J}(S, A)
\end{align*}
\paragraph{} We augment this conventional cost with an additional cost: the information processing cost. We argue that the brain has limited computational resources and it is costly to carry out complex computations required for information processing. We formalize this intuition through the framework of information bottleneck \citep{tishby_information_2000}. In this case, we are still performing optimal sensorimotor control to minimize an accuracy cost; however, limitations on the extent of  information processing that is possible is modeled as a constraint on the mutual information between sensory states $s$ and motor actions $a$ \citep{lai_human_2024}.
The optimization then takes takes the following form:
\begin{align*}
    \min_{p(a|s)} \mathcal{J}(S, A) \quad \textrm{s.t.} \quad I(S;A) \leq C \quad \textrm{and}  \quad  \sum_{a} p(a|s) = 1
\end{align*}
where we have also included a normalization constraint on our stochastic policy that the sum of the probabilities for all possible actions in a given state should be equal to 1.
The solution to this constrained optimization problem is well-established \citep{cover1991information, tishby_information_2000, tishby2010information,parush2011dopaminergic, yeung2012first}. A detailed, step-by-step derivation has not yet been provided. Here, we provide such derivation. 

\paragraph{}The total accuracy cost $\mathcal{J}(S, A)$ is the expected cost over all possible actions and states, i.e. given by summing accuracy costs for each state/action pair, weighted by the probability of each case occurring:
$$\mathcal{J}(S, A) = \sum_s \sum_a p(a,s) J(s,a)$$ 
The mutual information \(I(S;A)\) is defined as
\begin{align*}
    I(S;A) = \sum_s \sum_a p(s, a) \log \frac{p(s, a)}{p(s)p(a)} = \sum_s \sum_a p(s)p(a|s) \log \frac{p(a|s)}{p(a)}
\end{align*}
In order to solve the constrained optimization problem, we rewrite the function in Lagrangian form:
\begin{align*}
    \mathcal{L} = \mathcal{J}(S, A) + \beta (I(S;A) - C) + \nu \left(\sum_{a} p(a|s) - 1\right)
\end{align*}
The full Lagrangian function incorporating the mutual information term becomes,
\begin{align*}
    \mathcal{L} = \mathcal{J}(S, A) + \beta \left(\sum_s \sum_a p(s)p(a|s) \log \frac{p(a|s)}{p(a)} - C\right) + \nu \left(\sum_{a}p(a|s) - 1\right)
\end{align*}
\paragraph{} The optimal solution to this Lagrangian is determined by setting its partial derivative with respect to $p(a|s)$ at each $a_i$ and $s_j$ state-action instance to zero,
\begin{align*}
    \frac{\partial \mathcal{L}}{\partial p(a_i|s_j)} = 0\textbf{  } \quad \forall \textbf{ } (a_i, s_j) \in {S} \times {A}
\end{align*}
\paragraph{} We will break the Lagrangian into three terms. We will derive their partial derivatives separately for a given $a_i$ and $s_j$ state-action instance:
\begin{align*}
    \frac{\partial \mathcal{L}}{\partial p(a_i|s_j)}  = 
    \underbrace{
    \frac{\partial \mathcal{J}(S, A)}{\partial p(a_i|s_j)}}_{(1)}
    + 
   \beta \underbrace{
   \frac{\partial}{\partial p(a_i|s_j)}  \left[\sum_s \sum_a p(s)p(a|s) \log \frac{p(a|s)}{p(a)} - C\right]}_{(2)}
    + 
    \nu \underbrace{
    \frac{ \partial}{\partial p(a_i|s_j)}\left[\sum_{a} p(a|s_j) - 1\right]}_{(3)}
\end{align*}
\paragraph{} Focusing first on the first term, the accuracy cost term, in the Lagrangian:
\begin{align*}
     \underbrace{\frac{\partial \mathcal{J}(S, A)}{\partial p(a_i|s_j)}}_{(1)} = & \frac{ \partial } {\partial p(a_i|s_j)} \biggr[ \sum_s \sum_a p(a,s) J(s,a)\biggr]\\
      = & \frac{ \partial } {\partial p(a_i|s_j)} \biggr[ p(a_i|s_j) p(s_j) J(s_j,a_i)\biggr] + \frac{ \partial } {\partial p(a_i|s_j)} \biggr[\sum_{s \neq s_j} \sum_{a \neq a_i} p(a|s) p(s) J(s,a)\biggr] \\ 
      = & \frac{ \partial} {\partial p(a_i|s_j)} \biggr[p(a_i|s_j) p(s_j) J(s_j,a_i) \biggr]+ 0\\ 
      = & p(s_j) J(s_j,a_i) 
\end{align*}
\paragraph{} Now focusing on the second term, the information cost term, in the Lagrangian:
\begin{align*}
   \underbrace{
   \frac{\partial}{\partial p(a_i|s_j)}  \left[\sum_s \sum_a p(s)p(a|s) \log \frac{p(a|s)}{p(a)} - C\right]}_{(2)} =  \frac{\partial}{\partial p(a_i|s_j)}\left[\sum_s \sum_a p(s)p(a|s) \log \frac{p(a|s)}{p(a)}\right]= \frac{\partial I(S;A)}{\partial p(a_i|s_j)}
\end{align*}
We first expand the mutual information term, 
\begin{align*}
    I(S;A) = & \sum_s \sum_a p(s)p(a|s) \log \frac{p(a|s)}{p(a)} \\
    I(S;A) = & \underbrace{p(s_1) p(a_1|s_1) \log \frac{p(a_1|s_1)}{p(a_1)} + \dots }_{\sum_s p(s)p(a_1|s) \log \frac{p(a_1|s)}{p(a_1)}} + \underbrace{p(s_1) p(a_2|s_1) \log \frac{p(a_2|s_1)}{p(a_2)} + \dots}_{\sum_s p(s)p(a_2|s) \log \frac{p(a_2|s)}{p(a_2)}} + \dots \dots
\end{align*}
We will then have,
\begin{align*}
   \frac{\partial I(S;A)}{\partial p(a_i|s_j)} = & \frac{ \partial}{\partial p(a_i|s_j)}\left[ \underbrace{p(s_1) p(a_1|s_1) \log \frac{p(a_1|s_1)}{p(a_1)} + \dots }_{\sum_s p(s)p(a_1|s) \log \frac{p(a_1|s)}{p(a_1)}} + \underbrace{p(s_1) p(a_2|s_1) \log \frac{p(a_2|s_1)}{p(a_2)} + \dots}_{\sum_s p(s)p(a_2|s) \log \frac{p(a_2|s)}{p(a_2)}} + \dots \dots \right] \\
\end{align*}
Taken together, the information cost derivative $\frac{\partial I(S;A)}{\partial p(a_i|s_j)}$ is given by the following:
\begin{align*}
  \frac{\partial}{\partial p(a_i|s_j)}  \Biggl[ \Bigl[ p(s_j) p(a_i|s_j) \log \frac{p(a_i|s_j)}{p(a_i)} \Bigr]
   + \Bigl[ \sum_{s \neq  s_j} p(s) p(a_i|s) \log \frac{p(a_i|s)}{p(a_i)} \Bigr]
   + \Bigl[ \sum_{a \neq a_i} \sum_{s \neq  s_j} 
 p(s) p(a|s) \log \frac{p(a|s)}{p(a)} \Bigr] \Biggr]
\end{align*}
There are three separate terms in the information cost derivative $\frac{\partial I(S;A)}{\partial p(a_i|s_j)}$. We will solve each term separately. Let us first solve the first term,
\begin{align*}
       \frac{\partial}{\partial p(a_i|s_j)}&  \biggr[ \biggr. p(s_j)  p(a_i|s_j)  \log  \frac{p(a_i|s_j)}{p(a_i)}   \biggr. \biggr]\\ = & \left( \frac{ \partial }{\partial p(a_i|s_j)} \biggl[ p(s_j) p(a_i|s_j)\biggr] \right) \log \frac{p(a_i|s_j)}{p(a_i)} + p(s_j)p(a_i|s_j) \left( \frac{\partial  }{\partial p(a_i|s_j)} \biggl[ \log \frac{ p(a_i|s_j)}{p(a_i)}\biggr] \right)
       \\
        = & p(s_j)  \log \frac{p(a_i|s_j)}{p(a_i)} + p(s_j)p(a_i|s_j) \left( \frac{\partial}{\partial p(a_i|s_j)}  \biggr[\log \frac{ p(a_i|s_j)}{p(a_i)}\biggr] \right)
        \\
        = & p(s_j) \log \frac{p(a_i|s_j)}{p(a_i)} + p(s_j)p(a_i|s_j) \left( \frac{\partial}{\partial p(a_i|s_j)} \biggr[\log { p(a_i|s_j)}\biggr] - \frac{\partial }{\partial p(a_i|s_j)} \biggr[\log {p(a_i)}\biggr]\right)
\end{align*}
Continuing solving the above first term, we know that $p(a) = \sum_s p(s)p(a|s)$, where the probability of an action is obtained by summing the product of the probability of each state and the conditional probability of the action given that state over all possible states. After substituting $p(a)$ in the above first term, we obtain the following:
\begin{align*}
       \frac{\partial}{\partial p(a_i|s_j)} & \left[ p(s_j) p(a_i|s_j) \log \frac{p(a_i|s_j)}{p(a_i)} \right] \\ = & p(s_j)  \log \frac{p(a_i|s_j)}{p(a_i)} + p(s_j)p(a_i|s_j) \left( \frac{1}{p(a_i|s_j)} - \frac{\partial}{\partial p(a_i|s_j)} \biggr[ \log \sum_s p(s) p(a_i|s)\biggr]\right) \\
        = & p(s_j)  \log \frac{p(a_i|s_j)}{p(a_i)} + 
        p(s_j) - p(s_j)p(a_i|s_j)\frac{ \partial }{\partial p(a_i|s_j)} \biggr[\log \sum_s p(s) p(a_i|s)\biggr]
\end{align*}
Continuing with the derivative of the second term in the information cost term $\frac{\partial I(S;A)}{\partial p(a_i|s_j)}$,
\begin{align*}
        \frac{ \partial}{\partial p(a_i|s_j)} & \left[\sum_{s \neq  s_j} p(s) p(a_i|s) \log \frac{p(a_i|s)}{p(a_i)}\right] \\ 
        = & \frac{ \partial}{\partial p(a_i|s_j)} \left[\sum_{s \neq  s_j} p(s) p(a_i|s) \log \frac{p(a_i|s)}{\sum_s p(s) p(a_i|s)}\right] \\
         = & \frac{ \partial}{\partial p(a_i|s_j)} \left[\sum_{s \neq  s_j} p(s) p(a_i|s) \left( \log {p(a_i|s)} - \log {\sum_s p(s) p(a_i|s)} \right)\right] 
          \\
          = & \frac{ \partial }{\partial p(a_i|s_j)} \left[ \sum_{s \neq  s_j} p(s) p(a_i|s)  \log {p(a_i|s)} - {  \sum_{s \neq  s_j}p(s) p(a_i|s) \log {\sum_s p(s) p(a_i|s)} } \right]
          \\
          = & 0 - \frac{ \partial }{\partial p(a_i|s_j)}\left[\sum_{s \neq  s_j}p(s) p(a_i|s) \log {\sum_s p(s) p(a_i|s)} \right]
          \\
          = & - \sum_{s \neq  s_j}p(s) p(a_i|s) \frac{   \partial}{\partial p(a_i|s_j)} \biggr[ \log {\sum_s p(s) p(a_i|s)}  \biggr]
\end{align*}
The derivative of the third term in the information cost $\frac{\partial I(S;A)}{\partial p(a_i|s_j)}$ is trivial,
\begin{align*}
    \frac{ \partial }{\partial p(a_i|s_j)} \left[\sum_{a \neq a_i} \sum_{s \neq  s_j} 
 p(s) p(a|s) \log \frac{p(a|s)}{p(a)}\right] = 0
\end{align*}
 We combine the three terms we have separately derived for $\frac{\partial I(S;A)}{\partial p(a_i|s_j)}$ to obtain the following,
\begin{align*}
      \frac{\partial I(S;A)}{\partial p(a_i|s_j)} = &p(s_j)  \log \frac{p(a_i|s_j)}{p(a_i)} + 
        p(s_j) - p(s_j)p(a_i|s_j) \frac{  \partial }{\partial p(a_i|s_j)} \biggr[\log \sum_s p(s) p(a_i|s)\biggr] \\ & -\sum_{s \neq  s_j}p(s) p(a_i|s) \frac{   \partial}{\partial p(a_i|s_j)} \biggr[ \log {\sum_s p(s) p(a_i|s)}  \biggr]\\ 
       = & p(s_j)  \log \frac{p(a_i|s_j)}{p(a_i)} + 
        p(s_j) - \sum_{s}p(s) p(a_i|s)  \frac{ \partial }{\partial p(a_i|s_j)} \biggr[\log {\sum_s p(s) p(a_i|s)} \biggr]
\end{align*}
We focus on solving the last element in the above equation, 
\begin{align*}
    \sum_{s}p(s) p(a_i|s) & \frac{ \partial }{\partial p(a_i|s_j)} \biggr[\log {\sum_s p(s) p(a_i|s)} \biggr] \\
    = & \sum_{s}p(s) p(a_i|s) \left( \frac{1}{{\sum_s p(s) p(a_i|s)}} \frac{\partial }{\partial p(a_i|s_j)} \biggr[{\sum_s p(s) p(a_i|s)}\biggr]\right) \\
    = & \left( \sum_{s}p(s) p(a_i|s)   \frac{1}{{\sum_s p(s) p(a_i|s)}} \right) \left( \frac{\partial }{\partial p(a_i|s_j)}  \biggr[{\sum_{s \neq s_j} p(s) p(a_i|s)} +  p(s_j) p(a_i|s_j)\biggr]\right) \\
    = & \frac{\partial }{\partial p(a_i|s_j)}  \biggr[{\sum_{s \neq s_j} p(s) p(a_i|s)} +  p(s_j) p(a_i|s_j)\biggr]  \\
    = & \frac{\partial }{\partial p(a_i|s_j)} \biggr[{\sum_{s \neq s_j} p(s) p(a_i|s)}\biggr] + \frac{ \partial }{\partial p(a_i|s_j)} \biggr[p(s_j) p(a_i|s_j)\biggr] \\
    = & 0 + \frac{ \partial }{\partial p(a_i|s_j)} \biggr[p(s_j) p(a_i|s_j)\biggr]  \\
    = & p(s_j)
\end{align*}
Beautifully, everything is cancelled out and we have the following for the second term in the Lagrangian:
\begin{align*}
    \underbrace{
   \frac{\partial}{\partial p(a_i|s_j)}  \left[\sum_s \sum_a p(s)p(a|s) \log \frac{p(a|s)}{p(a)} - C\right]}_{(2)}= & p(s_j)  \log \frac{p(a_i|s_j)}{p(a_i)} + 
        p(s_j) - p(s_j) = p(s_j)  \log \frac{p(a_i|s_j)}{p(a_i)}
\end{align*}
\paragraph{} Let us focus on solving the third term, the normalization term, in the Lagrangian:
\begin{align*}
    & \underbrace{
    \frac{ \partial}{\partial p(a_i|s_j)}\left[\sum_{a} p(a|s_j) - 1\right]}_{(3)} \\ & \quad =  \frac{ \partial}{\partial p(a_i|s_j)}\left[p(a_i|s_j) + \sum_{a \neq a_i} p(a|s_j) - 1\right]
    \\ & \quad =  \frac{ \partial p(a_i|s_j)}{\partial p(a_i|s_j)}
    + \frac{ \partial}{\partial p(a_i|s_j)}\left[\sum_{a \neq a_i} p(a|s_j) - 1\right]
    \\ & \quad =  \frac{ \partial p(a_i|s_j)}{\partial p(a_i|s_j)}
    + 0 = 1 
\end{align*}
\paragraph{} We have solved each element in the Lagrangian separately and here we combine them, 
\begin{align*}
    \frac{\partial \mathcal{L}}{\partial p(a_i|s_j)}  = &
    \underbrace{
    \frac{\partial \mathcal{J}(S, A)}{\partial p(a_i|s_j)}}_{(1)}
    + 
   \beta \underbrace{
   \frac{\partial}{\partial p(a_i|s_j)}  \left[\sum_s \sum_a p(s)p(a|s) \log \frac{p(a|s)}{p(a)} - C\right]}_{(2)}
    + 
    \nu \underbrace{
    \frac{ \partial}{\partial p(a_i|s_j)}\left[\sum_{a} p(a|s_j) - 1\right]}_{(3)}  \\
  = & \underbrace{p(s) J(s,a) }_{(1)}
    + 
   \beta \underbrace{ p(s)  \log \frac{p(a|s)}{p(a)}}_{(2)}
    + 
    \nu \underbrace{1}_{(3)}  \\
     = &p(s) \left( J(s,a)
    + 
   \beta \log \frac{p(a|s)}{p(a)}
    + 
    \frac{\nu }{p(s)}\right)
\end{align*}
\paragraph{} The solution for the Lagrangian is given as $    \frac{\partial \mathcal{L}}{\partial p(a|s)} = 0$, and thus we have the following:
\begin{align*}
    p(s) \left( J(s,a)
    + 
   \beta \log \frac{p(a|s)}{p(a)}
    + 
    \frac{\nu }{p(s)}\right) & = 0 \\
    J(s,a)
    + 
   \beta \log \frac{p(a|s)}{p(a)}
    + 
    \frac{\nu }{p(s)} & = 0 \\
   \beta \log \frac{p(a|s)}{p(a)}
 & = - J(s,a) - \frac{\nu }{p(s)} \\
 \log \frac{p(a|s)}{p(a)}
 & = \frac{- J(s,a) - \frac{\nu }{p(s)}}{\beta}
\end{align*}
By simplifying the logarithmic term, the equation transforms into:
\begin{align*}
    \frac{p(a|s)}{p(a)} = & e^{\frac{- J(s,a) - \frac{\nu }{p(s)}}{\beta}} \\
    p(a|s) = & p(a) e^{{- \frac{1}{\beta}J(s,a) - \frac{\nu }{ \beta p(s)}}} \\
   p(a|s) = & \frac{p(a) e ^{{- \frac{1}{\beta}J(s,a)}}}{e^{\frac{\nu }{ \beta p(s)}}} 
\end{align*}
Moving $ {e^{\frac{\nu }{ \beta p(s)}}}$ to the left, we obtain, 
\begin{align*}
   p(a|s) {e^{\frac{\nu }{ \beta p(s)}}}  = & {p(a) e ^{{- \frac{1}{\beta}J(s,a)}}}
\end{align*}
We know that $\sum_{a} p(a|s) = 1$. Therefore, we have the following,
\begin{align*}
    \sum_a p(a|s) {e^{\frac{\nu }{ \beta p(s)}}} = & \sum_a {p(a) e ^{{- \frac{1}{\beta}J(s,a)}}}\\
    1 \cdot {e^{\frac{\nu }{ \beta p(s)}}} = & \sum_a {p(a) e ^{{- \frac{1}{\beta}J(s,a)}}}
    \\
    {e^{\frac{\nu }{ \beta p(s)}}} = & \sum_a {p(a) e ^{{- \frac{1}{\beta}J(s,a)}}}
\end{align*}
Putting $ {e^{\frac{\nu }{ \beta p(s)}}}$ back, we obtain:
\begin{align*}
    p(a|s) = & \frac{p(a) e ^{{- \frac{1}{\beta}J(s,a)}}}{\sum_a {p(a) e ^{{- \frac{1}{\beta}J(s,a)}}}} 
\end{align*}
The solution to our optimization problem can be also written as:
$$p(a|s) \propto p(a) e ^{{- \frac{1}{\beta}J(s,a)}}$$
\paragraph{} The mathematical derivation here can also be viewed as a  generalized version of the Blahut-Arimoto algorithm for rate distortion problems \citep{blahut1972computation, arimoto1972algorithm}.


\end{document}